\documentclass[twocolumn,english]{revtex4-1}
\usepackage[T1]{fontenc}
\usepackage[latin9]{inputenc}
\usepackage{amsmath}
\usepackage{amssymb}
\usepackage{graphicx}

\makeatletter
\usepackage{hyperref}

\makeatother

\usepackage{babel}
\begin{document}
\title{Real-space condensation of reciprocal active particles driven by spontaneous
symmetry breaking induced nonreciprocity}
\author{Wei-chen Guo}
\affiliation{Guangdong Provincial Key Laboratory of Quantum Engineering and Quantum
Materials, School of Physics and Telecommunication Engineering, South
China Normal University, Guangzhou 510006, China}
\affiliation{Guangdong-Hong Kong Joint Laboratory of Quantum Matter, South China
Normal University, Guangzhou 510006, China}
\author{Zuo Wang}
\affiliation{Guangdong Provincial Key Laboratory of Quantum Engineering and Quantum
Materials, School of Physics and Telecommunication Engineering, South
China Normal University, Guangzhou 510006, China}
\affiliation{Guangdong-Hong Kong Joint Laboratory of Quantum Matter, South China
Normal University, Guangzhou 510006, China}
\author{Pei-fang Wu}
\affiliation{Guangdong Provincial Key Laboratory of Quantum Engineering and Quantum
Materials, School of Physics and Telecommunication Engineering, South
China Normal University, Guangzhou 510006, China}
\affiliation{Guangdong-Hong Kong Joint Laboratory of Quantum Matter, South China
Normal University, Guangzhou 510006, China}
\author{Li-jun Lang}
\email{ljlang@scnu.edu.cn}

\affiliation{Guangdong Provincial Key Laboratory of Quantum Engineering and Quantum
Materials, School of Physics and Telecommunication Engineering, South
China Normal University, Guangzhou 510006, China}
\affiliation{Guangdong-Hong Kong Joint Laboratory of Quantum Matter, South China
Normal University, Guangzhou 510006, China}
\author{Bao-quan Ai}
\email{aibq@scnu.edu.cn}

\affiliation{Guangdong Provincial Key Laboratory of Quantum Engineering and Quantum
Materials, School of Physics and Telecommunication Engineering, South
China Normal University, Guangzhou 510006, China}
\affiliation{Guangdong-Hong Kong Joint Laboratory of Quantum Matter, South China
Normal University, Guangzhou 510006, China}
\author{Liang He}
\email{liang.he@scnu.edu.cn}

\affiliation{Guangdong Provincial Key Laboratory of Quantum Engineering and Quantum
Materials, School of Physics and Telecommunication Engineering, South
China Normal University, Guangzhou 510006, China}
\affiliation{Guangdong-Hong Kong Joint Laboratory of Quantum Matter, South China
Normal University, Guangzhou 510006, China}
\begin{abstract}
We investigate the steady-state and dynamical properties of a reciprocal
many-body system consisting of self-propelled active particles with
local alignment interactions that exists within a fan-shaped neighborhood
of each particle. We find that the nonreciprocity can emerge in this
reciprocal system once the spontaneous symmetry breaking is present,
and the effective description of the system assumes a non-Hermitian
structure that directly originates from the emergent nonreciprocity.
This emergent nonreciprocity can impose strong influences on the properties
the system. In particular, it can even drive a real-space condensation
of active particles. Our findings pave the way for identifying a new
class of physics in reciprocal systems that is driven by the emergent
nonreciprocity.
\end{abstract}
\maketitle
\emph{Introduction.}---The exceptional role played by the nonreciprocity
in various many-body systems has recently attracted much attention
in a wide range of fields in physics, ranging from condensed matter
physics \citep{Jiangbin_Gong_PRL_2019,Metelmann_PRX_2015,Scheibner_Nat_Phys_2020,Helbig_Nat_Phys_2020,Coulais_Nature_2017,Liew_PRB_2021,Lang_PRB_2021,Wang_PRB_2022,Ka_PRB_2022,Lv_PRB_2022,Bartolo_PRL_2022},
over statistical physics \citep{Fruchart_Nature_2021,Zhihong_You_PNAS_2020,Kryuchkov_SoftMatter_2018,Saha_PRX_2020,Chen_PRB_2021},
to biological physics \citep{Kondo_Science_2010,Lowen_PRE_2011,Vicsek_NJP_2017,Lowen_EPL_2021},
etc. Well-known examples include novel critical phenomena in nonreciprocal
phase transitions \citep{Fruchart_Nature_2021}, rich collective dynamical
behaviors in the predator-prey systems \citep{Lotka_JPhyschem_1910,Zhdankin_PRE_2010,Angelani_PRL_2012,Vicsek_NJP_2017,Lowen_EPL_2021},
Turing patterns in the activator-inhibitor systems \citep{Nakao_Nat_Phys_2010,Kosek_PRL_1995,Kundu_PRE_2021},
and non-Hermitian skin effects in the lattice models with nonreciprocal
hopping \citep{Yao_PRL_2018,Gong_PRX_2018}.

In fact, one can notice that the nonreciprocity giving rise to the
rich nonreciprocal physics in these scenarios is directly built-in
by, for instance, the nonreciprocal interactions between different
types of agents. While fundamental laws of physics, such as Newton\textquoteright s
third law, are usually reciprocal, hence also the physical systems
governed by them. In these regards, it seems that the rich nonreciprocal
physics are irrelevant for ubiquitous reciprocal systems. However,
in the spirit of ``\emph{more is different}'' pioneered by P. W.
Anderson \citep{Anderson_Science_1972}, one could actually still
expect that the nonreciprocity can emerge in reciprocal many-body
systems under certain circumstances. This thus gives rise to a novel
scenario for physics associated with the nonreciprocity, and raises
the fundamental question of the existence and the physical consequences
of the emergent nonreciprocity in reciprocal many-body systems.

\begin{figure}
\noindent \begin{centering}
\includegraphics[width=3.3in]{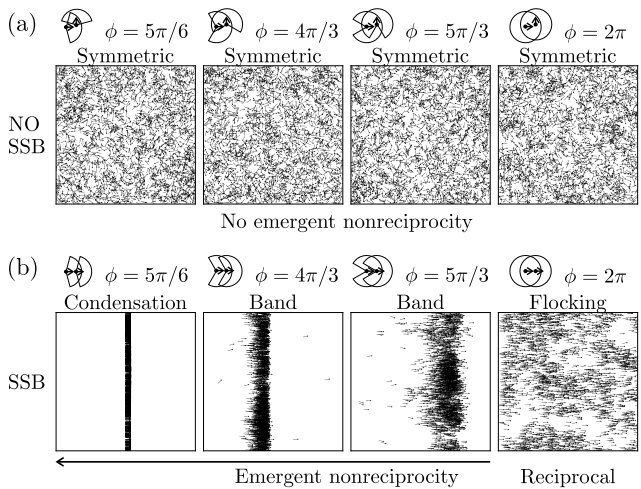}
\par\end{centering}
\caption{\label{fig:Illustration}(a) Typical steady-state configurations of
the system at a high noise level $\eta=3\slash4$ above the flocking
transition, i.e., no SSB happens in the system. Although the limited
view angle $\phi$ of active particles allows the existence of a transient
front-back nonreciprocity between certain pairs of adjacent particles
(illustrated at the top), there is no overall nonreciprocity present
in the symmetric disordered phase of the system. (b) Typical steady-state
configurations of the system at a low noise level $\eta=1\slash10$
below the flocking transition. The emergent nonreciprocity can be
induced by the SSB and impose strong influences on the system. In
particular, it can even drive a real-space condensation of active
particles (the leftmost panel). See text for more details.}
\end{figure}

\begin{figure}
\noindent \begin{centering}
\includegraphics[width=3.3in]{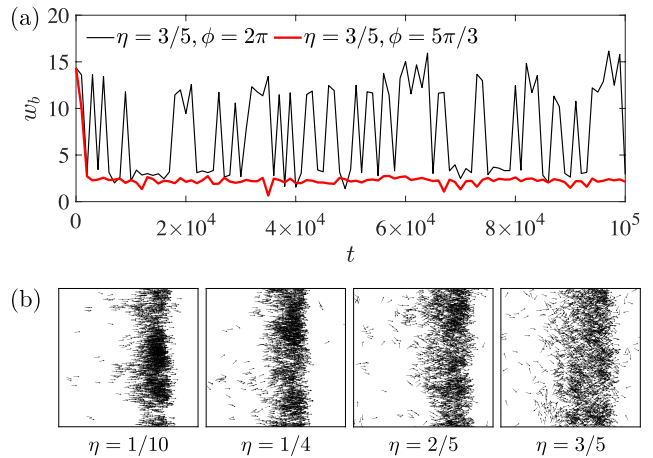}
\par\end{centering}
\caption{\label{fig:Trajectory}(a) Typical single trajectory time evolution
of the traveling band width $w_{b}$ for the system ($N=10^{4}$,
$L=50$) at the noise level $\eta=3/5$ slightly below the flocking
transition. The black and red curves correspond to the view angle
$\phi=2\pi,\,5\pi/3$, respectively. The temporal fluctuation of the
traveling band width $w_{b}$ is clearly suppressed by the emergent
nonreciprocity with $\phi=5\pi/3$. (b) Typical steady-state configurations
of the system ($N=2500$, $L=25$) with a fixed limited view angle
$\phi=5\pi/3$ at different noise levels $(\eta=3/5,2/5,1/4,1/10)$.
The emergent nonreciprocity with the limited view angle drives the
formation of the traveling band in a considerably large interval of
the noise level. See text for more details.}
\end{figure}

\begin{figure}
\noindent \begin{centering}
\includegraphics[width=3.3in]{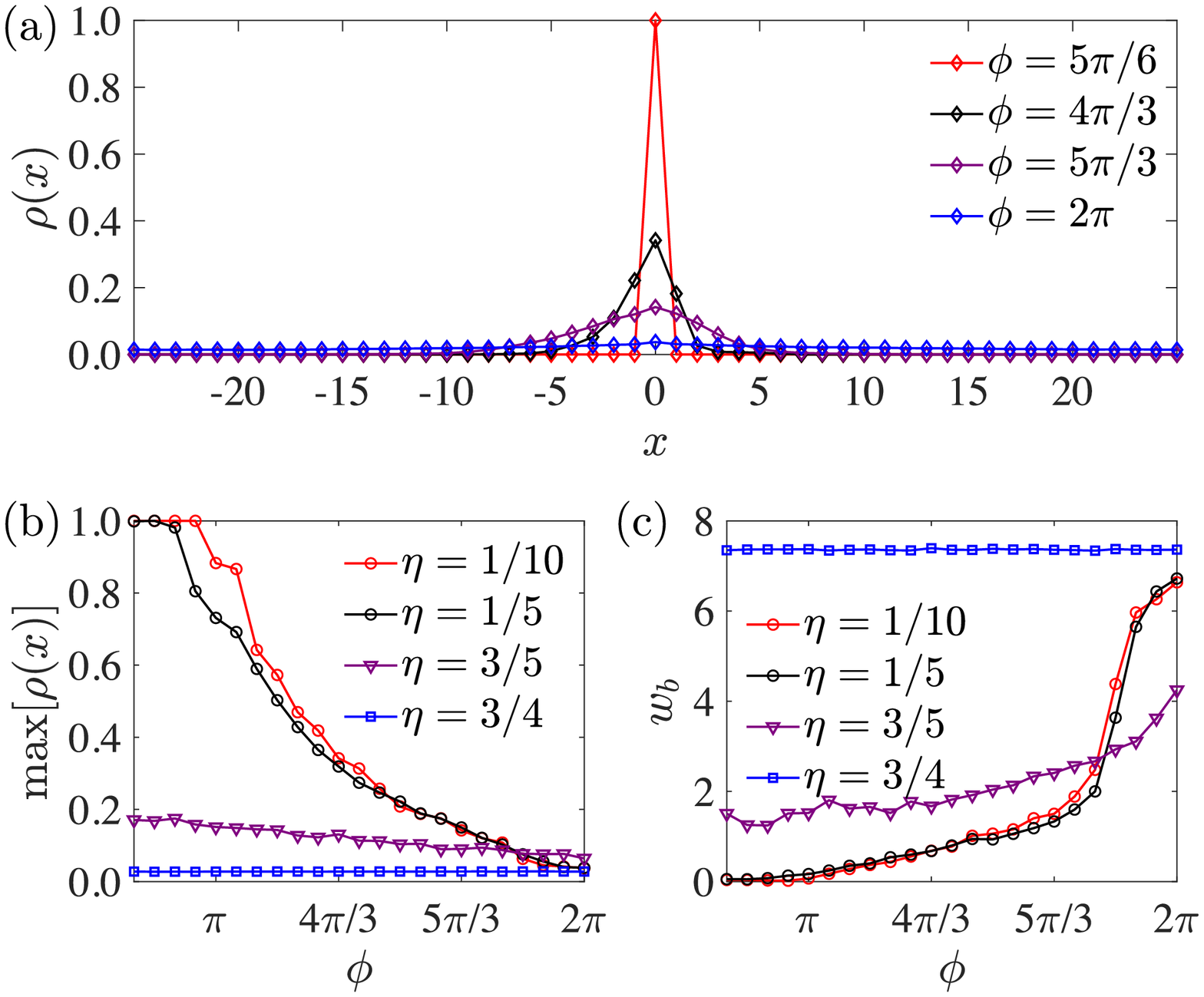}
\par\end{centering}
\caption{\label{fig:Main}(a) Normalized density distribution $\rho(x)$ along
the direction of the collective velocity with different view angles
at a low noise level $\eta=1\slash10$ below the flocking transition.
The peak positions of all $\rho(x)$ have been shifted to $x=0$ for
better comparison. Real-space condensation of active particles can
be clearly noticed with the view angle $\phi$ slightly below $\pi$
which corresponds to the strongest emergent nonreciprocity. (b) The
view angle $\phi$ dependence of the peak value of the normalized
density distribution $\rho(x)$ (denoted as $\max[\rho(x)]$) at different
noise levels. At the low noise levels $\eta=1/10,\,1/5$, the peak
value $\max[\rho(x)]$ reaches $1$ with a small enough view angle
$\phi<\pi$, indicating the formation of the real-space condensation
of active particles. (c) The view angle $\phi$ dependence of the
traveling band width $w_{b}$ at different noise levels. At the low
noise levels $\eta=1/10,\,1/5,\,3/5$ below the flocking transition,
the traveling band width $w_{b}$ assumes the smaller value with the
smaller $\phi$. While at the high noise level $\eta=3/4$ above the
flocking transition, $w_{b}$ is essentially independent of the view
angle $\phi$. This manifests that the emergent nonreciprocity is
induced by the SSB. See text for more details.}
\end{figure}

In this work, we address this question for a two-dimensional (2D)
reciprocal system consisting of self-propelled active particles with
local alignment interactions that exists within a fan-shaped neighborhood
of each particle {[}see Eq.~(\ref{eq:dynamics}) and the top of Fig.~\ref{fig:Illustration}(a){]}.
We find that the nonreciprocity can emerge in this reciprocal many-body
system and impose strong influences on the properties of the system.
More specifically, we find the following. (i) Emergence of nonreciprocity
that is induced by the spontaneous symmetry breaking (SSB). At the
low noise level below the flocking transition, the system spontaneously
breaks the rotational symmetry and induces the nonreciprocity between
the active particles located relatively in the front and the ones
located relatively in the back along the direction of the collective
velocity {[}see the top of Fig.~\ref{fig:Illustration}(b){]}. In
this case, the effective description of the system assumes a non-Hermitian
structure that directly originates from the emergent nonreciprocity
{[}see Eq.~(\ref{eq:Fokker-Planck}){]}. (ii) Emergent nonreciprocity
driven traveling band formation and real-space condensation of active
particles. The traveling band is very elusive in similar related active
matter systems \citep{Chate_PRL_2015,Chate_PRE_2015,Liebchen_PRL_2017,Chate_PRL_2021_Topological,Chate_PRL_2021_Quenched},
here we find that the emergent nonreciprocity can enhance the formation
of the traveling band in a considerably much larger parameter regime
(see Fig.~\ref{fig:Trajectory}). Even more remarkably, at low and
intermediate noise levels, we find that a strong enough emergent nonreciprocity
could even drive a real-space condensation of active particles along
the direction of the collective velocity {[}see Fig.~\ref{fig:Main}
and the leftmost panel of Fig.~\ref{fig:Illustration}(b){]}.

\emph{System and model.}---The system under study consists of $N$
self-propelled active particles moving in the 2D space of linear size
$L$ under influences from environmental fluctuations. Each particle
interacts with other particles via a local alignment interaction within
its fan-shaped neighborhood {[}see the top of Fig.~\ref{fig:Illustration}(a){]}.
The dynamics of the system is modeled by a modified Vicsek model with
extrinsic noise (also known as ``vectorial'' noise) \citep{Chate_PRE_2008,Chate_PRL_2004,Vicsek_PRL_1995,Toner_Ann_Phys_2005,Vicsek_Phys_Rep_2012,Volpe_RMP_2016,Chate_Annu_Rev_2020},
where the active particles move synchronously at discrete time steps
$\Delta t$ with a constant speed $v$ along the directions specified
by

\begin{equation}
\theta_{j}(t+\Delta t)=\arg\sum_{k\in U_{j}(r,\phi)}(e^{i\theta_{k}(t)}+\eta e^{i\xi_{j}(t)}).\label{eq:dynamics}
\end{equation}
Here, for the $j$th active particle at time $t$, $\theta_{j}(t)$
denotes its direction of motion in the 2D space, $U_{j}(r,\phi)$
denotes its fan-shaped neighborhood with $r$ being the radius of
the active particle's scope and $\phi$ being the view angle. The
environmental fluctuations are modeled by the random noise term $\eta\xi_{j}(t)$,
where $\xi_{j}(t)$ a random variable whose value is uniformly distributed
within the interval $[-\pi,\pi]$ and $\eta\in[0,1]$ is the level
of the extrinsic noise. The numerical results presented in this work
are obtained by directly simulating the dynamical Eq.~(\ref{eq:dynamics})
with periodic boundary condition imposed. We set $r=1$ and $\Delta t=1$,
hence the distance and time are measured in the unit of $r$ and $\Delta t$,
respectively. If not specified in text, the total number of particles
$N=2500$, the linear size $L=25$, the speed $v=1\slash2$, and $10^{2}$
stochastic trajectories are used to perform ensemble average.

\emph{Emergent nonreciprocity and non-Hermitian effective description
in the presence of the SSB.}---From the form of Eq.~(\ref{eq:dynamics})
we notice that irrespective of the value of the view angle $\phi$,
each particle satisfies the same form of the dynamical equation, reflecting
that it imposes influences on the other particles in the same manner,
indicating that the system is reciprocal. This is in sharp contrast
to the dynamical equations for systems with the built-in nonreciprocity,
for instance, the predator-prey systems \citep{Lotka_JPhyschem_1910,Zhdankin_PRE_2010,Angelani_PRL_2012,Vicsek_NJP_2017,Lowen_EPL_2021},
the activator-inhibitor systems \citep{Nakao_Nat_Phys_2010,Kosek_PRL_1995,Kundu_PRE_2021},
the leader-follower systems \citep{Vicsek_Nature_2010,Zhou_EPL_2015,Pearce_PRE_2016},
etc., where different types of agents satisfy different forms of dynamical
equations, reflecting the fact that they impose influences on each
other in intrinsically different ways. However, we also notice that
once the view angle $\phi<2\pi$, a transient front-back nonreciprocity
between any two different particles can actually exist if one particle
lies in the view of the other one but not vice versa {[}see for instance
the top of Fig.~\ref{fig:Illustration}(a){]}.

As one can see from Fig.~\ref{fig:Illustration}(a), which shows
the typical steady-state configurations of this reciprocal system
at a relatively high noise level $\eta$ above the flocking transition,
there is essentially no difference among configurations with different
view angles. This indicates that the transient front-back nonreciprocity
does not affect the properties of this reciprocal system in the symmetric
disordered phase. Indeed, since the direction of motion of each particle
changes randomly all the time in the symmetric disordered phase, there
is no overall nonreciprocity presented in the system.

But the SSB of the rotational symmetry can cause the system to develop
the flocking phase at a relatively low noise level \citep{Chate_PRE_2008,Chate_PRL_2004,Vicsek_PRL_1995,Toner_Ann_Phys_2005,Vicsek_Phys_Rep_2012,Volpe_RMP_2016,Chate_Annu_Rev_2020},
where the macroscopic number of particles moving along almost the
same direction of the collective velocity $\mathbf{v}_{c}\equiv N^{-1}\sum_{j=1}^{N}\mathbf{v}_{j}$
(with $\mathbf{v}_{j}$ being the velocity of the $j$th particle).
In this case, with the limited view angle $\phi<2\pi$, the influences
of the active particles located relatively in the front (``front-particles'')
along the direction of $\mathbf{v}_{c}$ on the ones located relatively
at the back (``back-particles'') are stronger than the influences
of ``back-particles'' on ``front-particles'' {[}see for instance
the top of Fig.~\ref{fig:Illustration}(b){]}. This ``influence-imbalance''
thus indicates that in the presence of the SSB, the overall nonreciprocity
emerges in the system between these two ``types'' of active particles.
Since the smaller view angle $\phi$ results in the larger ``influence-imbalance''
hence the stronger nonreciprocity, one naturally expects that with
a small enough view angle $\phi$, the emergent nonreciprocity could
strongly affect the physical behavior of the system. Indeed, as one
can see from Fig.~\ref{fig:Illustration}(b), which shows the typical
steady-state configurations of this reciprocal system at a low noise
level $\eta$ below the flocking transition, decreasing the view angle
$\phi$ (hence increasing the strength of the emergent nonreciprocity)
clearly changes the steady-state configurations of the system, i.e.,
from the typical flocking configuration with no additional structure,
over the traveling band that is very elusive in the conventional Vicsek
model \citep{Chate_PRL_2015,Chate_PRE_2015,Liebchen_PRL_2017,Chate_PRL_2021_Topological,Chate_PRL_2021_Quenched},
eventually to a remarkable real-space condensation line {[}see the
leftmost panel of Fig.~\ref{fig:Illustration}(b){]}.

To investigate how this emergent nonreciprocity affects the system,
let us first use a simple zero-dimensional effective model to describe
the exchange process between the ``front-particles'' and ``back-particles''
in the system. This effective model consists of two groups of particles
labeled by $F$ and $B$, whose dynamics is determined by two dynamical
rules $F\overset{\lambda_{FB}}{\rightarrow}B$ and $F\overset{\lambda_{BF}}{\leftarrow}B$,
where $F$ and $B$ denote any ``front-particle'' and ``back-particle'',
respectively. $\lambda_{FB}$ ($\lambda_{BF}$) is the rate, i.e.,
the probability within a unit time, for a single ``front-particle''
(``back-particle'') to change into a ``back-particle'' (``front-particle'').
The state of this model system at time $t$ is determined by the probability
distribution $P(n_{B},n_{F};t)$ that denotes the probability for
the model system in a configuration with $n_{F}$ ``front-particles''
and $n_{B}$ ``back-particles''. With the two dynamical rules, the
Fokker--Planck equation that determines the time evolution of the
probability distribution $P(n_{B},n_{F};t)$ can be straightforwardly
obtained, which assumes the explicit form

\begin{align}
 & \partial_{t}P(n_{B},n_{F};t)\label{eq:Fokker-Planck}\\
= & \left(e^{-\partial_{n_{B}}},e^{-\partial_{n_{F}}}\right)\Lambda\left(\begin{array}{c}
e^{\partial_{n_{B}}}n_{B}\\
e^{\partial_{n_{F}}}n_{F}
\end{array}\right)P(n_{B},n_{F};t),\nonumber 
\end{align}
with $\Lambda\equiv\left(\begin{array}{cc}
-\lambda_{BF} & \lambda_{FB}\\
\lambda_{BF} & -\lambda_{FB}
\end{array}\right).$ One immediately notices that the matrix $\Lambda$ that appears in
this Fokker--Planck equation is a non-Hermitian matrix in general
as long as $\lambda_{BF}\neq\lambda_{FB}$. Indeed, in the presence
of the SSB, the corresponding $\lambda_{BF}$ is expected to be larger
than $\lambda_{FB}$ with the limited view angle $\phi<2\pi$ due
to the emergent nonreciprocity. One thus expects that major effects
of the emergent nonreciprocity on the system can be captured by the
effective non-Hermitian description Eq.~(\ref{eq:Fokker-Planck}).
The steady-state solution of Eq.~(\ref{eq:Fokker-Planck}) can be
obtained analytically. The average number of ``front-particles''
$\langle n_{F}\rangle$ and its particle number fluctuations $\Delta n_{F}\equiv\langle n_{F}^{2}\rangle-\langle n_{F}\rangle^{2}$
in the steady-state assume the form 
\begin{equation}
\langle n_{F}\rangle=N\frac{\lambda_{BF}}{\lambda_{FB}+\lambda_{BF}},\,\Delta n_{F}=N\frac{\lambda_{BF}\lambda_{FB}}{(\lambda_{FB}+\lambda_{BF})^{2}}.\label{eq:steady_state_solution}
\end{equation}
From the above steady-state properties of the ``front-particles'',
one can see that a large imbalance between $\lambda_{BF}$ and $\lambda_{FB}$
causes more particles to concentrate in the front, and at the same
time reduces the corresponding particle number fluctuations. This
thus suggests that the physical effects of the emergent nonreciprocity
are to compress the particle number distribution along the direction
of the collective velocity which thus facilitates the formation of
the traveling band, and to make the band more stable by reducing the
particle number fluctuations within it. As we shall see in the following,
the system can form stable traveling band in a large parameter regime.
Even more remarkably, it can manifest a novel real-space condensation
of active particles at the large enough emergent nonreciprocity.

\emph{Emergent nonreciprocity driven traveling band formation and
real-space condensation.}---In the conventional Vicsek model (corresponding
to Eq.~(\ref{eq:dynamics}) with the full view angle $\phi=2\pi$),
the traveling band is quite elusive and can exist only when the noise
level of the system is within a very narrow interval slightly below
the flocking transition \citep{Chate_PRL_2015,Chate_PRE_2015,Liebchen_PRL_2017,Chate_PRL_2021_Topological,Chate_PRL_2021_Quenched}.
Its dynamics features alternate disintegration and reconstruction
of the band structure \citep{Chate_PRE_2008}, giving rise to large
particle number fluctuations in the band region that clearly manifests
its fragileness. From the above study on the non-Hermitian effective
description Eq.~(\ref{eq:Fokker-Planck}) of the system in the presence
of the SSB, we have seen that the emergent nonreciprocity with the
limited view angle $\phi<2\pi$ can assume the effect of reducing
the particle number fluctuations within the band. This thus indicates
that the stability of the traveling band can be substantially enhanced
by the emergent nonreciprocity.

Fig.~\ref{fig:Trajectory}(a) shows the single trajectory time evolution
of the traveling band width $w_{b}$ for the systems with $\phi=2\pi,\,5\pi/3$.
Here, the traveling band width $w_{b}$ is defined according to $w_{b}\equiv\sqrt{\langle x^{2}\rangle-\langle x\rangle^{2}}$
with $\langle x^{2}\rangle=\int\rho(x)x^{2}dx$ and $\langle x\rangle=\int\rho(x)xdx$,
where $\rho(x)$ is the normalized density distribution along the
direction of the collective velocity, i.e., $\rho(x)=N^{-1}\int n(x,y)dy$,
with $x$ and $y$ being the coordinates along and perpendicular to
the collective velocity, respectively, and $n(x,y)$ being the particle
number density at $(x,y)$. As one can see from Fig.~\ref{fig:Trajectory}(a),
the temporal fluctuation of the traveling band width $w_{b}$ is clearly
suppressed by the emergent nonreciprocity with the limited view angle
$\phi=5\pi/3$, thus making the traveling band more stable. Consequently,
one can naturally expect that in the presence of the emergent nonreciprocity,
the traveling band can also exists in a larger parameter regime. For
instance, as already shown clearly in Fig.~\ref{fig:Illustration}(b),
the system does not support the traveling band at the low noise level
$\eta=1/10$ in the absence of the emergent nonreciprocity at $\phi=2\pi$
{[}see the rightmost panel of Fig.~\ref{fig:Illustration}(b){]},
while upon decreasing the view angle $\phi$, the emergent nonreciprocity
drives the formation of the traveling band {[}see the two middle panels
of Fig.~\ref{fig:Illustration}(b){]}. Further simulations of the
system with a fixed limited view angle $\phi=5\pi/3$ at different
noise levels $(\eta=3/5,2/5,1/4,1/10)$ show in Fig.~\ref{fig:Trajectory}(b)
that the emergent nonreciprocity can indeed drive the formation of
the traveling band in a considerably large interval of the noise level
(with fixed $\phi=5\pi/3$, the traveling band is generally observed
at $1\slash10\leqslant\eta\leqslant3/5$).

From the study on the non-Hermitian effective description Eq.~(\ref{eq:Fokker-Planck})
of the system in the presence of the SSB, we also notice that more
particles shall concentrate in the front if the imbalance between
$\lambda_{BF}$ and $\lambda_{FB}$ becomes larger. Particularly,
in the extreme case where $\lambda_{FB}$ is negligibly small compared
to $\lambda_{BF}$, all the particles shall concentrate in the front.
This indicates that they shall organize themselves in such a way that
share the same position along the direction of the collective velocity,
i.e., form a real-space condensation along this direction. Fig.~\ref{fig:Main}(a)
shows the normalized density distribution $\rho(x)$ along the direction
of the collective velocity with different view angles at a fixed low
noise level $\eta=1/10$. One can see that as the view angle is decreased
from $2\pi$, the normalized density distribution $\rho(x)$ become
narrower. In particular, with a small enough view angle $\phi$ that
is slightly below $\pi$, the peak value of the normalized density
distribution $\rho(x)$, denoted as $\max[\rho(x)]$, reaches 1. This
indicates that all the active particles in the system condense at
the same position along the direction of the collective velocity {[}see
also the leftmost panel of Fig.~\ref{fig:Illustration}(b){]}. Indeed,
the ``front-particles'' can hardly impose any influences on the
``back-particles'' by the alignment interaction with $\phi<\pi$,
therefore $\lambda_{FB}$ is negligibly small compared to $\lambda_{BF}$
at low noise levels.

To further investigate the influence of the extrinsic noise on the
existence of the real-space condensation, we calculate the view angle
$\phi$ dependence of $\max[\rho(x)]$ at different noise levels.
As shown in Fig.~\ref{fig:Main}(b), one can see that real-space
condensation generally exists at low ($\eta=1/10$) and intermediate
($\eta=1/5$) noise levels when the view angle $\phi$ is decreased
below $\pi$. At the relatively high noise level $(\eta=3/5)$, no
real-space condensation exists even if the view angle $\phi$ is decreased
below $\pi$. This is due to the fact that although the ``front-particles''
can hardly impose any influences on the ``back-particles'' by the
alignment interaction with $\phi<\pi$, the strong environmental fluctuations
can still give a finite contribution to $\lambda_{FB}$.

Influences from environmental fluctuations are also reflected directly
in the view angle $\phi$ dependence of the traveling band width $w_{b}$
as shown in Fig.~\ref{fig:Main}(c). At low ($\eta=1/10$) and intermediate
($\eta=1/5$) noise levels, the emergent nonreciprocity induced by
the view angle $\phi$ below $\pi$ is strong enough to drive $w_{b}$
to decrease to zero, hence the formation of the real-space condensation.
At the relatively high noise level $(\eta=3/5)$ that is still below
the flocking transition, the density fluctuation caused by the large
extrinsic noise can prevent the traveling band width $w_{b}$ from
decreasing to zero even in the presence of the emergent nonreciprocity.
While at the high noise level $(\eta=3/4)$ above the flocking transition,
the change of the view angle $\phi$ does not impose any noticeable
effect on $w_{b}$, which is a clear manifestation of the fact that
the nonreciprocity is induced by the SSB.

Finally, we remark that the emergent nonreciprocity driven traveling
band formation and real-space condensation is quite robust against
choices of particle densities, finite size effects, and even the type
of the noise in the system (see Supplemental Material \citep{Supplemental_material}
for details). This naturally facilitates further direct experimental
observations of these emergent nonreciprocity driven physics.

\emph{Conclusions.}---Nonreciprocity can emerge in reciprocal systems
and crucially influences their physical properties, as the emergent
nonreciprocity in the reciprocal system of active particles with the
limited view angle reveals. The SSB of the rotational symmetry of
this system can induce the nonreciprocity between the active particles
located relatively in the front and the ones located relatively in
the back along the direction of the collective velocity. In particular,
this gives rise to a remarkable real-space condensation of active
particles that is absent in the corresponding reciprocal system without
the emergent nonreciprocity. Since the mechanism for the nonreciprocity
to emerge in reciprocal systems is quite general, we expect that there
exists a large class of emergent nonreciprocity driven new physics
in reciprocal systems to be identified. For instance, the flock of
active spins \citep{Solon_PRL_2013,Caussin_PRL_2015,Chatterjee_PRE_2020}
are expected to manifest similar real-space condensation driven by
the emergent nonreciprocity. Moreover, the SSB in many other reciprocal
active matter systems with their dynamics determined by intelligent
decision-making rules, such as the intelligent materials \citep{Kaspar_Nature_2021},
systems with herding dynamics \citep{Boguna_PRE_2015,Boguna_PRE_2017},
and systems with quorum sensing \citep{Yusufaly_PRE_2016_QuorumSensing,Aguilar_PRE_2021_QuorumSensing,Tobias_Nat_Commun_2018_QuorumSensing},
etc., might also induce the nonreciprocity and give rise to new physics.
We believe our findings and predictions will stimulate further theoretical
and experimental efforts in revealing the emergent nonreciprocity
driven new physics in various reciprocal many-body systems.

\emph{Note added.}---Upon completion of this manuscript, we became
aware of the work by Loos et al. \citep{Loos_arXiv_2022}, reporting
long-range order and directional defect propagation in a 2D XY model
with vision cone interactions that is associated with the zero speed
limit of the model investigated in this manuscript.
\begin{acknowledgments}
This work was supported by NSFC (Grants No.~11874017, No.~11904109,
No.~12075090), Guangdong Basic and Applied Basic Research Foundation
(Grant No.~2019A1515111101), Natural Science Foundation of Guangdong
Province (Grant No.~2022A1515010449), Science and Technology Program
of Guangzhou (Grant No.~2019050001), and START grant of South China
Normal University.
\end{acknowledgments}

\end{document}


\renewcommand{\theequation}{S-\arabic{equation}} \renewcommand{\thefigure}{S\arabic{figure}} \renewcommand{\thetable}{S\arabic{table}}
\title{Supplemental Material for ``Real-space condensation of reciprocal
active particles driven by spontaneous symmetry breaking induced nonreciprocity''}
\author{Wei-chen Guo}
\affiliation{Guangdong Provincial Key Laboratory of Quantum Engineering and Quantum
Materials, School of Physics and Telecommunication Engineering, South
China Normal University, Guangzhou 510006, China}
\affiliation{Guangdong-Hong Kong Joint Laboratory of Quantum Matter, South China
Normal University, Guangzhou 510006, China}
\author{Zuo Wang}
\affiliation{Guangdong Provincial Key Laboratory of Quantum Engineering and Quantum
Materials, School of Physics and Telecommunication Engineering, South
China Normal University, Guangzhou 510006, China}
\affiliation{Guangdong-Hong Kong Joint Laboratory of Quantum Matter, South China
Normal University, Guangzhou 510006, China}
\author{Pei-fang Wu}
\affiliation{Guangdong Provincial Key Laboratory of Quantum Engineering and Quantum
Materials, School of Physics and Telecommunication Engineering, South
China Normal University, Guangzhou 510006, China}
\affiliation{Guangdong-Hong Kong Joint Laboratory of Quantum Matter, South China
Normal University, Guangzhou 510006, China}
\author{Li-jun Lang}
\affiliation{Guangdong Provincial Key Laboratory of Quantum Engineering and Quantum
Materials, School of Physics and Telecommunication Engineering, South
China Normal University, Guangzhou 510006, China}
\affiliation{Guangdong-Hong Kong Joint Laboratory of Quantum Matter, South China
Normal University, Guangzhou 510006, China}
\author{Bao-quan Ai}
\affiliation{Guangdong Provincial Key Laboratory of Quantum Engineering and Quantum
Materials, School of Physics and Telecommunication Engineering, South
China Normal University, Guangzhou 510006, China}
\affiliation{Guangdong-Hong Kong Joint Laboratory of Quantum Matter, South China
Normal University, Guangzhou 510006, China}
\author{Liang He}
\affiliation{Guangdong Provincial Key Laboratory of Quantum Engineering and Quantum
Materials, School of Physics and Telecommunication Engineering, South
China Normal University, Guangzhou 510006, China}
\affiliation{Guangdong-Hong Kong Joint Laboratory of Quantum Matter, South China
Normal University, Guangzhou 510006, China}
\maketitle

\section*{Influences of particle density, system size, and the type of noise}

\begin{figure}
\noindent \begin{centering}
\includegraphics[width=3.3in]{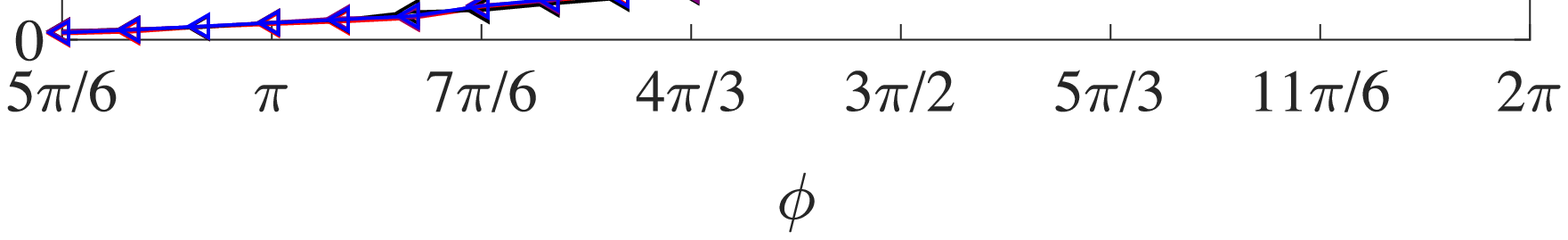}
\par\end{centering}
\caption{\label{fig:Other-properties}The view angle $\phi$ dependence of
the traveling band width $w_{b}$ in different cases. These results
show that the particle density, system size, and the type of noise
do not assume strong influences on the physical effects of the emergent
nonreciprocity. See text for more details.}
\end{figure}

For the sake of completeness, it is worth noting that several other
properties in this nonequilibrium complex system might also affect
its collective behaviors \citep{Chate_PRE_2008,Chate_PRL_2004,Vicsek_PRL_1995,Toner_Ann_Phys_2005,Vicsek_Phys_Rep_2012,Volpe_RMP_2016,Chate_Annu_Rev_2020}.
The particle density $\bar{n}=N\slash L^{2}$ is one that cannot be
ignored. A high particle density $\bar{n}$ makes the system relatively
robust against influences from environmental fluctuations, but as
one can see from the red and black lines in Fig.~\ref{fig:Other-properties},
reducing the density $\bar{n}$ within a certain range do not prevent
the real-space condensation from existing. The purple line in Fig.~\ref{fig:Other-properties}
indicates that finite size effects do not impose strong influences
on the physical effects of the emergent nonreciprocity, neither. Besides,
when the emergent nonreciprocity is strong enough to driven the real-space
condensation, we find that larger systems (e.g., $N=10^{4}$) can
accommodate several traveling ``lines'' at the same time without
forming regular wave trains, which is similar to the cases associated
with the well-established traveling band \citep{Chate_PRE_2008}.
Furthermore, concerning the fluctuations, the extrinsic noise characterizes
influences from environmental fluctuations and assumes that active
particles are totally willing to be aligned with their neighbors.
There is another type of noise in the Vicsek model, namely, the intrinsic
noise (also known as ``angular'' noise) \citep{Vicsek_PRL_1995},
where the dynamics of active particles follows $\theta_{j}^{t+\Delta t}=(\arg\sum_{k\in U_{j}(r,\phi)}e^{i\theta_{k}^{t}})+\tilde{\eta}\xi_{j}^{t}$,
with $\tilde{\eta}$ being the intrinsic noise strength characterizing
influences from the ``free will'' of the active particles \citep{Vicsek_PRL_1995}.
As one can see from the blue line in Fig.~\ref{fig:Other-properties},
the emergent nonreciprocity driven traveling band formation and real-space
condensation can also be found in this case. These results show that
the particle density, system size, and the type of noise do not assume
strong influences on the physical effects of the emergent nonreciprocity.